# An ANOVA Test for Parameter Estimability using Data Cloning with Application to Statistical Inference for Dynamic Systems

Corresponding Author:

David Campbell
Department of Statistics and Actuarial Science
Simon Fraser University
Surrey, BC V3T 0A3
Canada
dac5@sfu.ca

Subhash Lele
Department of Mathematical and Statistical Sciences
University of Alberta
Edmonton, AB T6G 2G1
Canada
slele@ualberta.ca

Abstract:

Models for complex systems are often built with more parameters than can be uniquely identified by available data. Because of the variety of causes, identifying a lack of parameter identifiability typically requires mathematical manipulation of models, monte carlo simulations, and examination of the Fisher Information Matrix.  A simple test for parameter estimability is introduced, using Data Cloning, a Markov Chain Monte Carlo based algorithm.  Together, Data cloning and the ANOVA based test determine if the model parameters are estimable and if so, determine their maximum likelihood estimates and provide asymptotic standard errors. When not all model parameters are estimable, the Data Cloning results and the ANOVA test can be used to determine estimable parameter combinations or infer identifiability problems in the model structure. The method is illustrated using three different real data systems that are known to be difficult to analyze.

1.INTRODUCTION

The ability of dynamic system models to succinctly describe complex behavior with a few but readily interpretable parameters has led to their wide spread popularity. Dynamic system models often try to model all the inherent behavior of the underlying process but the observations do not always have adequate information to support such complexity, leading to parameter inestimability. The innovation of this paper is the introduction of a statistical test for parameter estimability based on Data Cloning (DC) (Lele, Dennis and Lutscher 2007). The test uses the result that for inestimable parameters, the posterior distribution converges asymptotically to a non-degenerate distribution (Lele, Nadeem and Schumland 2010), corresponding to a version of the prior distribution truncated on the subspace where the likelihood attains its maxima. On the other hand, if the parameters are estimable, the posterior converges to a degenerate distribution concentrated on the maximum likelihood estimator (MLE), where DC provides both the MLE and asymptotic standard error thereof. Even if the original parameters are not estimable, a function of these parameters may be estimable and the proposed test can be used to judge the estimability of such a function without re-running DC.

While the proposed test is applicable anywhere that a maximum likelihood estimator is appropriate, this paper focuses on dynamic systems applications because they are often over parameterized and it is typically difficult to asses estimability. To demonstrate the proposed test we therefore consider three challenging real life dynamic systems whose analysis using currently available methods is known to be extremely difficult and at times misleading. Although the data cloning method as a general tool for inference in hierarchical models is now reasonably known, its application and relevance to inference for dynamic systems has not been tested in real life situations of this complexity. Consequently an additional goal of this paper is to outline the DC implementation details subject to a variety of real world challenges.

We start with a discussion of the various definitions of identifiability and estimability in section 2. Section 3 discusses the method of data cloning with emphasis on the ANOVA based estimability test. In section 4, we consider three case studies and section 5 discusses extensions and some additional details.

2: ESTIMABILITY AND IDENTIFIABILITY IN DYNAMIC SYSTEM MODELS

We outline three forms of non-identifiability or inestimability. A more comprehensive description can be found in Wu et al. (2008).

*1) Structural Identifiability*: This concerns with whether the parameters are uniquely identifiable given that the correct model is used and the process is observed in continuous time without error. Determining structural identifiability involves transformations of variables, implicit functions or differential algebra before data are available (Ljung and Glad 1994, Xia and Moog 2003).

*2) Practical Identifiability*: Despite being structurally identifiable, noisy data may produce parameters with infinite or otherwise impractical confidence intervals. Practical identifiability is subjective in that it may depend on the question being asked and the method used for producing confidence intervals. Assessing practical identifiability is commonly performed using artificial data sets based on a-priori parameter values to explore the sensitivity of the MLE to data uncertainty (Yao et al. 2003 and Anh et al. 2006).

*3) Statistical Estimability*: Even if the process is structurally and practically identifiable, the time points at which the system is observed or the particular data set obtained may lead to more than one parameter value where the likelihood function is maximized. Statistical inestimability may result in rank deficiencies in the Fisher Information Matrix (FIM) or extreme parameter correlations (Rodriguez-Fernandez, Mendes and Banga 2006, Schittkowski 2007, Sulieman et al. 2009). Even without numerical instabilities in assessing matrix rank deficiencies and correlations, these diagnostic tests are not foolproof.

If the set of parameters over which the likelihood reaches its maximum involves separated modes, within each mode the FIM may be full rank. Exploration of the profile likelihood (Raue et al 2009) has also been used to study statistical estimability. However when lack of identifiability is determined through the profile likelihood, no insights into potentially identifiable functions of the parameters are available. It is important to note that parameters can only be statistically estimable if they are also structurally identifiable, yet parameters may be structurally and practically identifiable without being statistically estimable. This paper focuses on statistical estimability and we use the terms identifiability and estimability interchangeably to mean statistical estimability and/or structural identifiability.

3: DATA CLONING AND A TEST FOR ESTIMABILITY

In this section, we introduce model notation, outline the method of Data Cloning, and introduce the ANOVA test for estimability.

Consider a general system of M coupled differential equations, with vector valued state equations governed by parameters $\varphi$:

$$\frac{dx(t)}{dt} = f(x(t); \varphi), \tag{1}$$

that describe the rate of change of the system states with respect to time, $t$, as a function of a vector of M system states, $x(t)$. In practice, only a subset of the system states are observable at discrete time points, $t \in \{t_1,...,t_n\}$ subject to the measurement error distribution with parameter $\psi$. Typically (1) has no analytic solution and $x(t, \varphi, x_0)$ must be solved numerically based on the initial condition $x_0$. The observations $Y$, follow the distribution $Y \mid x(t, \varphi, x_0), \psi \sim h(Y; x_0, \varphi, \psi)$. The likelihood function for $\theta = [\varphi, \psi, x_0]$ from the n data points, $y_{(n)} = (y_1, y_2, ..., y_n)$ can be written as:

$$L(\theta; y_{(n)}) = \prod_{i=1}^{n} h(y_i; X = x[t_i, \varphi, x_0], \psi).$$

Evaluation of the likelihood function involves solving the underlying ODEs for fixed $\varphi$ and $x_0$, and integrating over the values of $x(t,\varphi,x_0)$ with respect to its measure dX. For Lipschitz continuous deterministic f(.), a unique $x(t,\varphi,x_0)$ exists. In the standard Bayesian approach, with prior distribution $\pi(\theta)$, the posterior distribution is given by $\pi(\theta | y_{(n)}) = \frac{L(\theta; y_{(n)})\pi(\theta)}{C(y_{(n)})}$ where $C(y_{(n)}) = \int L(\theta; y_{(n)})\pi(\theta)d\theta$ is the normalizing constant.

3.1 Data Cloning

Consider a hypothetical situation where the experiment resulting in observations $y_{(n)}$ is repeated independently by K different individuals and by chance all individuals obtain the identical set of observations $y_{(n)}$. Denote these data by $y^{(K)} = (y_{(n)}, y_{(n)}, ... y_{(n)})$. The posterior distribution of $\theta$ conditional on $y^{(K)}$ is $\pi(\theta | y^{(K)}) = \frac{[L(\theta; y_{(n)})]^K \pi(\theta)}{C(K; y_{(n)})}$ where $C(K, y_{(n)}) = \int [L(\theta; y_{(n)})]^K \pi(\theta)d\theta$. It follows from the asymptotic behavior of the posterior that under regularity conditions, for large K, $\pi(\theta | y^{(K)})$ is approximately Normal with mean $\hat{\theta}$ and K times its variance is the asymptotic variance of the MLE, i.e. k times the inverse FIM (Walker, 1969; Lele et al. 2010). In data cloning, we create this hypothetical experiment by constructing the K-cloned data set, $y^{(K)}$, by copying the original data K times. We pretend that these data were obtained from K independent experiments and use MCMC to generate random samples from $\pi(\theta | y^{(K)})$ to obtain the posterior mean (MLE) and variance (k times inverse FIM) (Lele, et al. 2007; Lele, et al. 2010).

3.2 Test for Parameter Estimability

To design an estimability test, consider what happens to the posterior distribution as we increase K when using a unimodal proper prior. If the parameters are estimable, the set $\Omega = \left\{ \theta : \left[L(\theta; y_{(n)})\right]^K = \sup_{\theta \in \Theta} \left[L(\theta; y_{(n)})\right]^K \right\}$ contains a single point whereas if the parameters are not estimable, $\Omega$ consists of more than a single point. Lele et al. (2010) show that as $K \to \infty$, the DC posterior converges to a degenerate distribution truncated on $\Omega$. Consequently, if the parameters are estimable, the posterior variance converges to 0 as $K \to \infty$. If $\Omega$ contains a manifold in the parameter space, the posterior variance converges to some positive number and the posterior location depends on the prior. If $\Omega$ contains disjoint points, then DC will converge to locally degenerate distributions where the different priors will emphasize one mode over another. Consequently, the DC posterior mean will be prior dependent if parameters are inestimable.

This provides the conceptual basis for the proposed diagnostic test for parameter estimability: *Select different priors and check if the posterior mean is sensitive to this choice. If they are nearly identical for vastly different priors, the parameters are estimable.*

We can also use this test to diagnose estimability of a specified function of the parameters. Note that the sensitivity of the posterior mean to the prior might arise for two reasons: because the parameters are inestimable, and/or the number of clones is insufficient. If one can show that the number of clones is unlikely to be the reason for prior sensitivity, then sensitivity to the prior is likely due to inestimability. This diagnostic test can therefore be formulated as two sequential ANOVA tests. For expositional simplicity, let us start with a single parameter case. Let $\hat{\phi}_{k,p}$ denote the posterior mean with $k$ clones and the $p^{th}$ prior in the linear model $\hat{\phi}_{k,p} = \beta + \lambda_{k,p} + \varepsilon_{k,p}$ subject to the usual ANOVA constraint $\sum_k \lambda_{k,p} = 0$.

The first test checks the assumptions of the second test. Testing that all $\lambda_{k,p} = 0$ is a test for significance of the cloning effect; i.e. if sufficient clones have been used. Specifically, we test the following:

H$_{0(\text{clone})}$: There is no difference between point estimates when changing clones, i.e: $\lambda_{k,p} = 0, \forall k$. In other words an adequate number of clones has been used so that all chains have converged to the same point estimates regardless of the number of clones applied.

H$_{1(clone)}$: At least one point estimate differs from those arising from different amounts of cloning, i.e.: not all $\lambda_{k,p} = 0$. In this situation additional clones are required because differences in $\hat{\phi}_{k,p}$ based on the level of cloning are beyond what we expect if H$_{0(clone)}$ is true.

The underlying statistical model is based on testing if posterior means are converging to fixed locations with changes in k.
If the null hypothesis is rejected, runs with low k should be discarded and new runs with higher k should be included until the number of clones is no longer causing an impact. The multi-parameter case can be handled by using MANOVA or several single parameter ANOVAs because different amounts of information are available from the likelihood for different parameters.

When no significant effect due to cloning is found, we test the estimability hypothesis based on the linear model: $\hat{\phi}_{k,p} = \alpha + \delta_p + \varepsilon_{k,p}$ subject to the usual ANOVA constraint $\sum_p \delta_p = 0$. Testing the null hypothesis that all $\delta_p = 0$, assesses estimability by checking if there is a significant change in the DC posterior mean due to changes in the prior. Specifically we test the hypotheses:

H$_{0(prior)}$: There is no difference between point estimates when changing priors, i.e: $\delta_p = 0, \forall p$ and therefore the parameter is identifiable.

H$_{1(prior)}$: At least one point estimates differs from those arising from different prior specifications, i.e: not all $\delta_p = 0$, and therefore the parameter is not identifiable.

This second test is based on Monte Carlo variability where H$_{0(prior)}$ states that all chains have converged to the same location regardless of the prior distribution chosen. H$_{1(prior)}$ says that the differences in $\hat{\phi}_{k,p}$ are larger than can be attributed to Monte Carlo variation alone. Since we have already established that there is no significant cloning effect, we pool across clones for replications and use them to obtain an estimate of the Monte Carlo variation. The type 2 error probability for this test can be made arbitrarily small by choosing distant priors scattered around the parameter space, and thereby increasing the prior effect should it exist.

Note that using the full set of posterior draws as replicates, given the large sample size, the tests become highly sensitive to minute differences that arise due to Monte Carlo error rather than real differences in the mean. Consequently we consider the ANOVA tests using the posterior means and not full set of posterior samples.

Again, the multi-parameter case can be handled by several single parameter ANOVAs since interest is in testing estimability of individual parameters. It is important to note that each of the MCMC runs must be run to within chain convergence, and burn-in must be assessed and discarded using, for example, a Raftery Lewis (1992) diagnostic before performing the hypothesis tests.

If the hypothesis tests are not rejected for any of the parameters, then there is not enough evidence to suggests that the data cloning runs are targeting different likelihood values. These DC samples can then be combined in order to utilize information contained in all the intermittent results. The combined MLE averages across the K clone settings and $n_p$ priors using weights $w_{k,p} = {N_{k,p}}\big/{\sum_k \sum_{p=1}^{n_p} N_{k,p}}$ accounting for the $N_{k,p}$ MCMC draws: $\hat{\phi} = \sum_k \sum_{p=1}^{n_p} w_{k,p} \hat{\phi}_{k,p}$.

The combined posterior variance estimate uses a k-scaled weighted average variance $\text{var}(\hat{\phi}) = \sum_k \sum_{p=1}^{n_p} w_{k,p} k \, \text{var}(\hat{\phi}_{k,p})$.

4: DYNAMIC SYSTEMS AND DATA CLONING

We consider three dynamic systems to illustrate the ANOVA estimability insights, detection of estimable parameter combinations, and provide DC implementation details. The first example analyzes nylon production data to show how to produce MLE and interval estimates of identifiable parameters while diagnosing inestimable parameters. The second example analyzes the Dow Chemical Company Differential Algebraic Equation system. This system is well known to have more parameters than can be estimated, yet there is a debate as to which parameters are estimable. With this system we infer uniquely identifiable parameter combinations. The third example is an epidemiological model with a mixture of discrete and continuous parameters. Parameter estimability in this context is somewhat different because the FIM is not defined and profile likelihood methods are required in conjunction with DC. In that example, we emphasize non-standard implementation details when the parameters are both discrete and continuous.

4.1 Locally Inestimable Model

To understand the dynamics involved in melt-phase nylon reactions, in a heated reactor amine ($A$) and carboxyl ($C$) were combined forming polyamine links ($L$) and water ($W$) (Zheng, McAuley, Marchildon and Yao 2005). Simultaneously in the backward reaction, water decomposed $L$ back into $A$ and $C$. Due to the heat in the reactor, some of the water dissipated as steam. To examine the reaction rates, $W$ in the form of steam, was bubbled through a molten nylon mixture to drive the competing chemical reactions towards equilibrium concentrations. Six replicate experiments were run with the level of input steam held high, then low, then returned back to its initial level. We consider the data and original model of Zheng et al (2005):

$$-\frac{dL}{dt} = \frac{dA}{dt} = \frac{dC}{dt} = -k_p\left(CA - \frac{LW}{K_a}\right),$$

$$\frac{dW}{dt} = k_p\left(CA - \frac{LW}{K_a}\right) - 24.3(W - W_{eq}).$$
(8)

Reaction rates $k_p$ and $K_a$ are determined by the 6 model parameters $\varphi = [k_{p0}, E, \alpha, \beta, K_{a0}, H]$, the theoretical equilibrium water level, $W_{eq}$, based on the input steam level, and the reactor temperature $T_i$ in the i$^{\text{th}}$ experiment through the additional relations:

$$k_p = k_{p0} 10^{-3} \exp\left(-\frac{E}{8.3145 \times 10^{-3}}\left[\frac{1}{T_i} - \frac{1}{549.15}\right]\right),$$

$$K_a = \frac{20.97(1 + g[\alpha,\beta,T_i]W_{eq})K_{a0}}{\exp\left(9.624 - \frac{3613}{T_i}\right)\exp\left(\frac{H}{8.3145 \times 10^{-3}}\left[\frac{1}{T_i} - \frac{1}{549.15}\right]\right)},$$
(9)

$$g(\alpha, \beta, T) = \exp(\alpha + 10^3 \beta / T).$$

Using the mass balance of the system, given any three of $A, C, L$ and $W$, the fourth can be algebraically computed. Only $A$ and $C$ are observable and are assumed to follow the error structure:

$$Y_A \sim N(S_A[\varphi, A, C, W], \sigma_A^2)$$
$$Y_C \sim N(S_C[\varphi, A, C, W], \sigma_C^2),$$
(10)

where $S_A(\varphi, A, C, W)$ and $S_C(\varphi, A, C, W)$ are the solutions to model (8) for A and C. Unlike Zheng et al (2005), we consider initial conditions $A(0), C(0), W(0)$ for each of the 6 experiments and observation error variances $\sigma_A^2$ and $\sigma_C^2$ as unknown bringing the total number of parameter to 26. Due to the choice of the temperatures in the experiment performed in Zheng et al (2005), parameters $(\alpha, \beta)$ are highly confounded and the likelihood is rife with ridges and plateaus. For example, for $\alpha \gg 10^3 |\beta|/T > 0$, $g(\alpha,\beta,T)$ is approximately constant over the range of $T_i$ at which the experiment was conducted. Consequently $\varphi$ reduces to 5 effective parameters with a local likelihood mode along a ridge in $\beta$. Additionally for $\alpha \ll -10^3 |\beta|/T$, we have $g(\alpha,\beta,T) \approx 0$ for all design values of $T_i$. In this region, $(\alpha,\beta)$ have no effect on the ODE reducing $\varphi$ to 4 effective parameters where the local likelihood plateaus over $(\alpha,\beta)$. In contrast, when $g(\alpha,\beta,T)>0$, such that $g(\alpha,\beta,T_i) \neq g(\alpha,\beta,T_j)$ when $T_i \neq T_j$, then $(\alpha,\beta)$ are in the basin of attraction of a global likelihood maximum. In this case $(\alpha,\beta)$ are locally identifiable, although they are extremely correlated (r>.995) due to the narrow range of values of $T_i$.

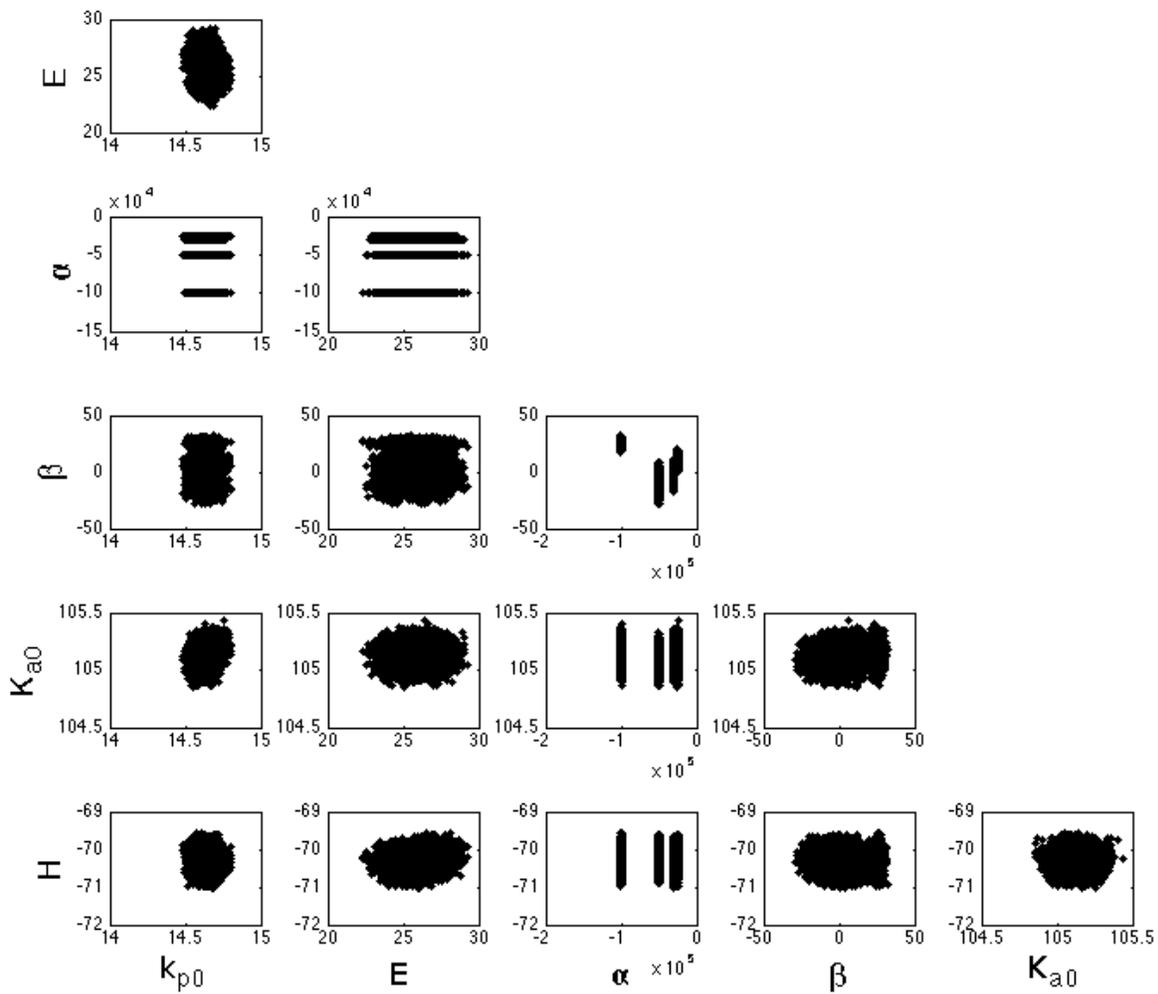

Figure 1. Bivariate plots of posterior samples for the Nylon parameters using 500 clones and four different prior distributions. The sensitivity to the choice of prior gives rise to appearance of gaps in the samples for $\alpha$ and $\beta$.

| parameter | | | | |
|---|---|---|---|---|
| $k_{p0}, E, K_{a0}, H$ | N(0,20) | N(-10,10) | N(-100,20) | N(50,10) |
| $\alpha$ | N(-1e6,1000) | N(-3e5,1000) | N(-2.5e4,100) | N(-5e4,200) |
| $\beta$ | N(25,2) | N(-3,4) | N(10,3) | N(-10,6) |

Table 1: Means and variances for the independent Gaussian prior distributions in the locally inestimable model of section 4.1.

For illustrative purposes, we use four different priors (listed in table 1) and restrict attention to the part of the parameter space where $g(\alpha,\beta,T) \approx 0$. Data cloning was performed with K={500,750,1000} and 25,000 posterior MCMC draws were kept after discarding burn-in. The cloning effect is not significant for any of the parameters, however the prior effect is significant for $\alpha$ and $\beta$ (p-values are given in table 2) indicating their inestimability. Figure 1 shows the matrix plot of bivariate posterior samples of $\varphi$, with 500 clones, from all 4 priors. The inestimability of $\alpha$ and $\beta$ and resulting sensitivity to different priors is demonstrated by posterior samples focused on well separated regions. No relationships between parameters are evident from the matrix plot, suggesting that the cause of inestimability is not due to parameter relationships, and instead, one should examine the role of $\alpha$ and $\beta$ in the model to diagnose that $g(\alpha,\beta,T) \approx 0$ in this region of the parameter space. On the other hand, samples for $[k_{p0},E,K_{a0},H]$ were deemed estimable by the ANOVA test and do not show such sensitivity. Point and interval estimates for estimable parameters are given in table 2. Because DC was performed focusing on the region where $g(\alpha,\beta,T) \approx 0$ the parameter estimates obtained from this analysis coincide with those from the simpler model where $g(\alpha,\beta,T)=0$ and therefore, $\varphi=[k_{p0},E,K_{a0},H]$. The role of diagnostic plots and re-parameterization is further explored in the next example.

|  | $k_{p0}$ | E | $\alpha$ | $\beta$ | $K_{a0}$ | H |
|---|---|---|---|---|---|---|
| Point and 95% interval | 14.6±2.1 | 26±45 | --- | --- | 105.1±3.3 | -70.3±9.9 |
| p-value for cloning effect | .28 | .45 | .57 | .29 | .46 | .67 |
| p-value for prior effect | .27 | .46 | 2x $10^{-18}$ | 5x $10^{-9}$ | .73 | .86 |

Table 2. Estimates and p-values for the nylon example.

4.2 Finding Estimable Parameter Combinations in the Presences of Structural non-Identifiability

We consider a model from the Dow Chemical Company describing catalyzed chemical reactions in an isothermal batch reactor (Biegler, Damiano and Blau 1986). The true chemical identities are proprietary and therefore disguised. Here we consider the full differential algebraic equation (DAE) model in its original specification, describing the behaviors of the 10 chemical components $[y_1,...y_{10}]$ in the presence of a catalyst $[Q^+]$ via 6 differential components:

$$\frac{dy_1}{dt} = -k_2 y_8 y_2$$

$$\frac{dy_2}{dt} = -k_1 y_6 y_2 + k_3 y_{10} - k_2 y_8 y_2$$

$$\frac{dy_3}{dt} = -k_2 y_8 y_2 + k_1 y_6 y_4 - .5 k_3 y_9$$

$$\frac{dy_4}{dt} = -k_1 y_6 y_4 + .5 k_3 y_9 \tag{11}$$

$$\frac{dy_5}{dt} = k_1 y_6 y_2 - k_3 y_{10}$$

$$\frac{dy_6}{dt} = -k_1 (y_6 y_2 + y_6 y_4) + k_3 (y_{10} + .5 y_9)$$

and 4 algebraic components:

$$y_7 = -[Q^+] + y_6 + y_8 + y_9 + y_{10}$$

$$y_8 = \frac{\theta_8 y_1}{\theta_8 + y_7}$$

$$y_9 = \frac{\theta_9 y_3}{\theta_9 + y_7} \tag{12}$$

$$y_{10} = \frac{\theta_7 y_5}{\theta_7 + y_7}$$

where T is the temperature and the parameters to estimate are $\varphi = [k_{10}, k_{20}, k_{30}, E_1, E_2, E_3, \theta_7, \theta_8, \theta_9]$, some of which arise from temperature the dependencies:

$$k_1 = k_{10} exp\left(-E_1 \left[T^{-1} - 340.15^{-1}\right]\right)$$

$$k_2 = k_{20} exp\left(-E_2 \left[T^{-1} - 340.15^{-1}\right]\right) \tag{13}$$

$$k_3 = k_{30} exp\left(-E_3 \left[T^{-1} - 340.15^{-1}\right]\right)$$

The original data set has 3 experimental runs, each at a different fixed temperature with observations for $\{y_1, y_2, y_3, y_4\}$ at unevenly spaced times. For pedagogical purposes, following Wu et al (2011) we consider only the data from the lowest temperature setting; $T = 313.15$ Kelvin. Pairs of the parameters $(k_{10}, E_1), (k_{20}, E_2)$ and $(k_{30}, E_3)$ therefore lack structural identifiability because T is fixed in (13). A Gaussian likelihood centered on the numerical solution to the DAE system was used to perform DC. Following others (Biegler et al (1986) and Wu et al (2011)), we consider $x_0$ known. We used K = {50,000, 60,000, 80,000} for each of 5 different multivariate Gaussian priors with parameters given in table 3. The null hypothesis of no cloning effect is not rejected with p-values for each parameter all above 0.20. While the values of K may seem large, the p-values were not as decisive and in some cases the cloning effect was significant when smaller K was used. This suggests that some parameters are technically estimable, but very little information is available about them.

| $k_{10}$ | $k_{20}$ | $k_{30}$ | $E_1$ | $E_2$ | $E_3$ | $\theta_7$ | $\theta_8$ | $\theta_9$ |
|---|---|---|---|---|---|---|---|---|
| (2.5,1) | (1,1) | (.3,.5) | (1,1) | (50,20) | (1,1) | (1500,50) | (30,.5) | (800,50) |
| (1,10) | (1,10) | (1,10) | (1,10) | (1,100) | (1,10) | (500,25) | (7,1) | (150,25) |
| (.5,1) | (.5,1) | (.5,1) | (5,3) | (10,10) | (5,3) | (400,35) | (5,1) | (200,50) |
| (2,1) | (2,1) | (2,1) | (2,1) | (25,1) | (5,3) | (300,25) | (3,1) | (100,50) |

Table 3: Mean and Standard Deviation of the independent Gaussian Priors for the DOW model.

As the number of clones increases, the bivariate densities of $(k_{10}, E_1), (k_{20}, E_2)$ and $(k_{30}, E_3)$ degenerate to densities along non-linear maximum likelihood manifolds defined by (13). Consequently, the null hypotheses of no prior effects was rejected with p-values < .00001 for all 6 of these parameters. While, this lack of structural identifiability can be derived mathematically, an important advantage of DC and the ANOVA estimability test is the ability to assess estimability without any mathematical manipulations. Furthermore, if the relationships in (13) are not known, reasonable guesses can be made based on examination of the bivariate posterior plots.

After the transformations described in equation (13), remaining analysis focuses on estimability of the model given by (11) and (12) with resulting parameter vector $[k_1, k_2, k_3, \theta_7, \theta_8, \theta_9]$. No significant difference was found between posterior means due to changing the prior for $k_1$ (p-value .98), $k_2$ (p-value .64), $k_3$ (p-value .35), and $\theta_8$ (p-value .46). However, the null hypothesis of no prior effect was rejected for $\theta_7$ and $\theta_9$ (p-values <.005).

Figure 2 shows bivariate posterior samples for $[k_1, k_2, k_3, \theta_7, \theta_8, \theta_9]$ based on 80,000 clones and all 5 prior distributions. A strong linear relationship is evident between $k_3$ and $\theta_8$, but the lack of sensitivity to the prior implies that these parameters are statistically estimable despite their strong correlation. Gaps in the posterior plots with respect to $\theta_7$ and $\theta_9$ are due to sensitivity to the priors noted in the ANOVA test. Closer inspection of figure 2 reveals a strong relationship between $\theta_7$ and $\theta_9$, consequently the transformed variable $\theta_7 / \theta_9$ was created and was, in fact, found identifiable (p-value .98). The transformed variable also appears in figure 2, where gaps due to changing the prior are no longer present. In short, the 9 parameter model is far more complex than the data can justify when considering a single experimental run. We find 5 estimable parameter combinations: $\{k_1, k_2, k_3, \theta_8, (\theta_7 / \theta_9)\}$ where point and intervals estimates are given in table 4. The extremely wide confidence intervals are due to the weakly informative likelihood in the neighborhood of the MLE, which was reflected in the large number of clones needed. Using the same data, Wu et al (2011) also suggest that 5 parameters are estimable when conditioning on holding the remaining parameters fixed. However their optimal parameter subset includes $\{E_1, E_2, k_3, E_3, \theta_7\}$. Although $(k_{30}, E_3)$ are not structurally identifiable, methods like that of Wu et al (2011) are based on finding linear relationships whereas the proposed ANOVA model does not require specification of the nature of the relationship.

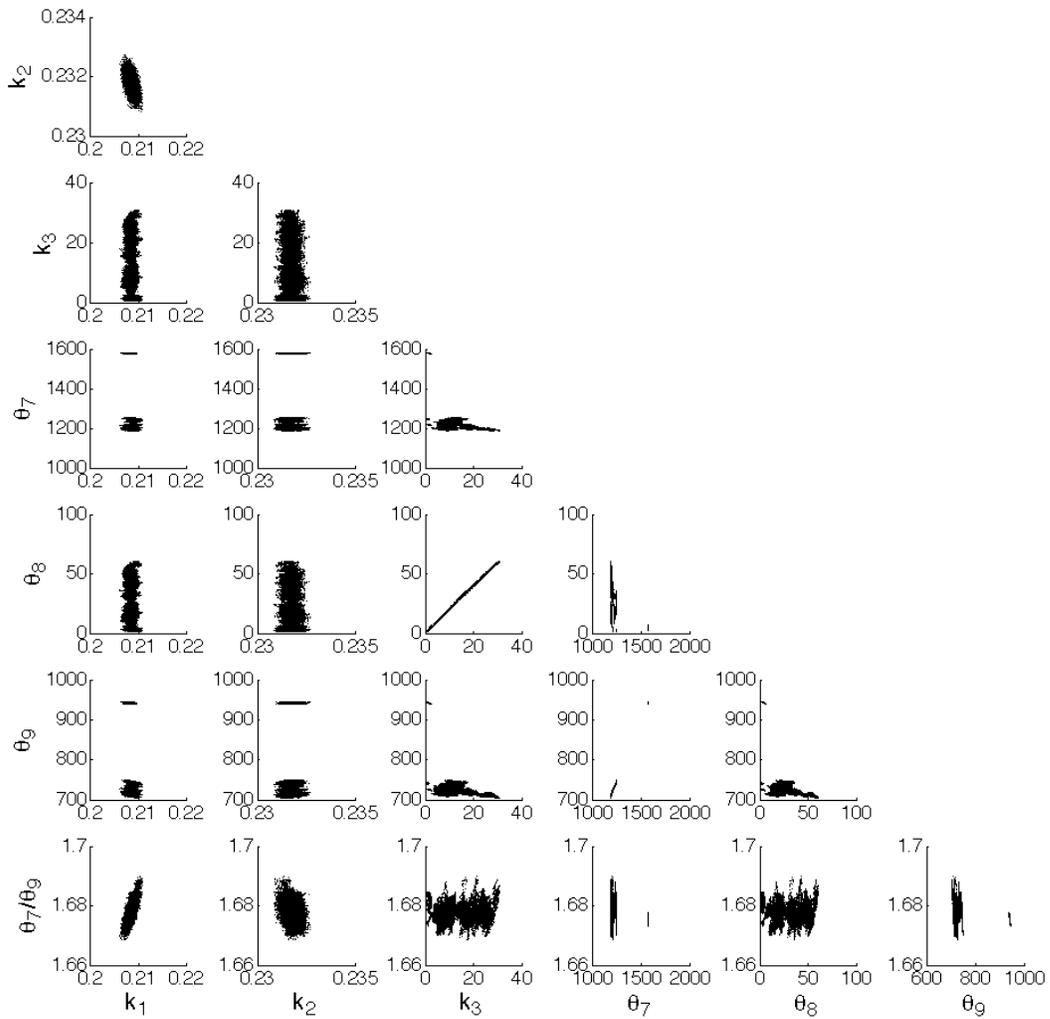

Figure 2: Matrix plot of 25,000 bivariate posterior samples from the DOW problem with 80,000 clones from each of 5 priors. The gaps in the samples with respect to $\theta_7$ and $\theta_9$ are due to sensitivity to the prior. The prior sensitivity is no longer present in plots of the transformed variable $\theta_7/\theta_9$ with respect to the other identifiable parameters.

| $k_1$ | $k_2$ | $k_3$ | $\theta_8$ | $(\theta_7/\theta_9)$ |
|---|---|---|---|---|
| 0.21±.27 | 0.23±.13 | (5.7±1145) x$10^4$ | 12±2290x$10^{-9}$ | 1.68±.70 |

Table 4. MLE and 95% intervals for the identifiable parameter combinations in the DOW example.

4.3: Mixtures of Discrete and Continuous Parameters

We consider an epidemiological model for a population consisting of susceptibles (S), infectious (I) and removed (R) individuals and infection is spread with parameters $\beta$ and $\alpha$ according to the dynamics:

$$\frac{d}{dt}S = -\beta SI, \qquad \frac{d}{dt}I = \beta SI - \alpha I, \qquad \frac{d}{dt}R = \alpha I \ . \quad (14)$$

Using this simple model, parameter estimates and identifiability analysis is performed using counts of daily deaths from second outbreak of the plague from June 19, 1666 until November 1, 1666 in the village of Eyam, UK, as recorded by the town gravedigger (Massad, Coutinho, Burattini and Lopez 2004). Because there is no recovery from the plague, the number of deaths up to time t corresponds to R(t) giving 136 observations of $Y_R$. The village had quarantined itself and hence one can assume that S(t)+ I(t)+R(t)=N is fixed at 261, the total population of the village, and that R(0)=0. Observations for S(t) and I(t) are not available, except at the end of the plague when the number of infected is 0 and the number of infected after the second to last death must therefore equal 1. We use DC to fit (14) to these data through the observed process:

$$Y_R \sim Poisson(R[t]) \quad (15)$$

where R(t) is obtained from the solution to equation (14) giving the likelihood:

$$L(\beta,\alpha,I(0)|Y_R) = \prod_{t=0}^{T} \frac{e^{-R(t)} R(t)^{Y_R}}{Y_R!}$$

$$= \prod_{t=0}^{T} \frac{\exp\left(-\alpha \int_0^t I(s)\,ds\right)\left(\alpha \int_0^t I(s)\,ds\right)^{Y_R}}{Y_R!}$$

. (16)

The 3 parameters; $[I(0), \alpha, \beta]$ are used along with the points $R(0) = 0$ and S(0)=N-R(0)-I(0), to numerically integrate (14) and compute the likelihood. Parameter $I(0)$ is discrete but $\alpha$ and $\beta$ are continuous, consequently the FIM is not defined. A straightforward maximization of the likelihood over this parameter space using DC, although technically possible, is difficult due to multimodality induced by the discrete nature of $I(0)$. As the number of clones increases so does the difficulty in obtaining a Markov chain which can efficiently travel between distant modes. On the other hand, if the value of $I(0)$ is fixed, the likelihood function for the remaining parameters $(\alpha, \beta)$ is well behaved as sometimes occurs with ODE systems (Wu, et al. 2008). Hence, we used DC to obtain the $\alpha$ and $\beta$ which maximize $L(\beta,\alpha|Y_R, I(0))$, the likelihood conditional on fixed values of $I(0)$.

Estimability and estimation are not straightforward so details are outlined in this section. In short, maximizing $L(\beta,\alpha|Y_R, I(0))$ for discrete values of $I(0)$ evaluates the profile likelihood for $I(0)$, profiling over $\alpha$ and $\beta$. To obtain the unconditional MLE we modify the Data Cloning Likelihood Ratio (DCLR) of Ponciano, Taper, Dennis and Lele (2009). These profile likelihood insights are used to find the unconditional MLE and confidence regions as described next.

1. *Conditional MLE:* Conditional on each value of $I^* = I(0) \in [1, 261]$, we ran DC with K={10, 100, 1000, 5000} copies of the data. We ran each chain in parallel with different starting values and each of 3 different sets of gamma priors for $\alpha$ and $\beta$ with parameters given in table 5. We applied the ANOVA identifiability diagnostic tests on the conditional likelihoods and found no evidence of cloning effect or inestimability in the conditional posteriors (p-values all > .30). Using these conditional posterior samples from K clones $\{\beta^{(I^*)}, \alpha^{(I^*)}\}_k$ we determined the conditional MLEs $\hat{\beta}^{(I^*)}$ and $\hat{\alpha}^{(I^*)}$.

| | | | |
|---|---|---|---|
| $\alpha$ | (1,1) | (3,2) | (2,1) |
| $\beta$ | (1,1) | (3,2) | (2,1) |

Table: 5, Gamma prior parameters for the plague dataset, parameterize so that if X~Gamma(a,b), then mean(X) = ab and var(X)=ab².

2. *Unconditional MLE:* We calculated the likelihood in (16) compared the values for each triad ($\hat{\beta}^{(I^*)}, \hat{\alpha}^{(I^*)}, I^*$) at a fixed number of clones to determine the overall MLE: ($\hat{\beta}^{(\hat{I})}, \hat{\alpha}^{(\hat{I})}, \hat{I}$). Note that unconditional estimability is assured if the MLE for $I(0)$ is unique and conditional on $I(0)$ parameters $\beta$ and $\alpha$ are estimable.

3. *Conditional and Profile Confidence Intervals:* For each candidate value of $I^*$, we modified the Ponciano et al (2009) DCLR test statistic to use the K cloned likelihood. The test statistic:

$$-\frac{2}{K}\log\left[\frac{L\left(\hat{\beta}^{(I^*)}, \hat{\alpha}^{(I^*)}, I^* | y^{(K)}\right)}{L\left(\hat{\beta}^{(\hat{I})}, \hat{\alpha}^{(\hat{I})}, \hat{I} | y^{(K)}\right)}\right] \sim \chi_1^2 ,$$

was used to determine values of $I^*$ which are contained in the profile interval region for $\hat{I}$. Any $I^*$ for which the test statistic exceeds the 95[th] percentile of a $\chi_1^2$ is excluded from the 95% profile confidence interval for $\hat{I}$. We use 1 degree of freedom because this DCLR is considering one parameter while profiling over the other 2 parameters. The resulting minimum 95% confidence set for $\hat{I}$ contains the values {4,5,6}. Confidence intervals conditional on $I^*$ are determined by multiplying the K cloned conditional posterior variance of $\{\beta^{(I^*)}, \alpha^{(I^*)}\}_k$ by K to get the conditional inverse FIM. Conditional point and interval estimates are given in table 6.

4. *Joint Confidence Regions:* We determined the 95% joint confidence regions for the triad ($\beta, \alpha, I$) by computing the similarly modified DCLR K-cloned likelihood ratio test statistic:

$$-\frac{2}{K}\log\left[\frac{L(\beta,\alpha,I|y^{(K)})}{L(\hat{\beta}^{(\hat{I})},\hat{\alpha}^{(\hat{I})},\hat{I}|y^{(K)})}\right] \sim \chi_3^2 \,, \qquad (17)$$

and determining the manifolds along which the likelihood ratio equals the 95$^{th}$ percentile of the $\chi_3^2$. Here we have 3 degrees of freedom because we are interested in joint inference on 3 parameters and are not profiling over any additional parameters. Contour finding can be performed using optimization methods or a grid search. For mathematical simplicity, we chose to use a stochastic search based on samples of ($\beta^{(I^*)}, \alpha^{(I^*)}, I^*$) to determine where the likelihood ratio approximately attains the 95$^{th}$ percentile of the $\chi_3^2$. Specifically, we exploit the asymptotic normality upon which the DC analysis is based and use the posterior mean and variance of the conditional posteriors of ($\beta^{(I^*)}, \alpha^{(I^*)}, I^*$). We then obtain a large sample from the cloned asymptotic conditional distributions by generating random normal random variables with appropriate mean and variances. Computing (17) for the sampled values of ($\beta^{(I^*)}, \alpha^{(I^*)}, I^*$) over a range of $I^*$ provides a fast parallelizable way of determining which of the samples are within the joint confidence region. Samples within the unconditional confidence regions are shown in Figure 3 where the disjoint regions highlight the bumpy nature of the likelihood surface with changes in *I*(0).

| Conditioning on $I(0)$ | $\beta$ | $\alpha$ |
|---|---|---|
| 4 | $5.7 \times 10^{-4}$ ($0.9 \times 10^{-4}$) | .087 (.002) |
| 5 | $6.2 \times 10^{-4}$ ($0.1 \times 10^{-4}$) | .096 (.003) |
| 6 | $6.8 \times 10^{-4}$ ($0.1 \times 10^{-4}$) | .108 (.003) |

Table 6. Point and standard error estimates for the Eyam plague data conditional on fixed $I(0)$.

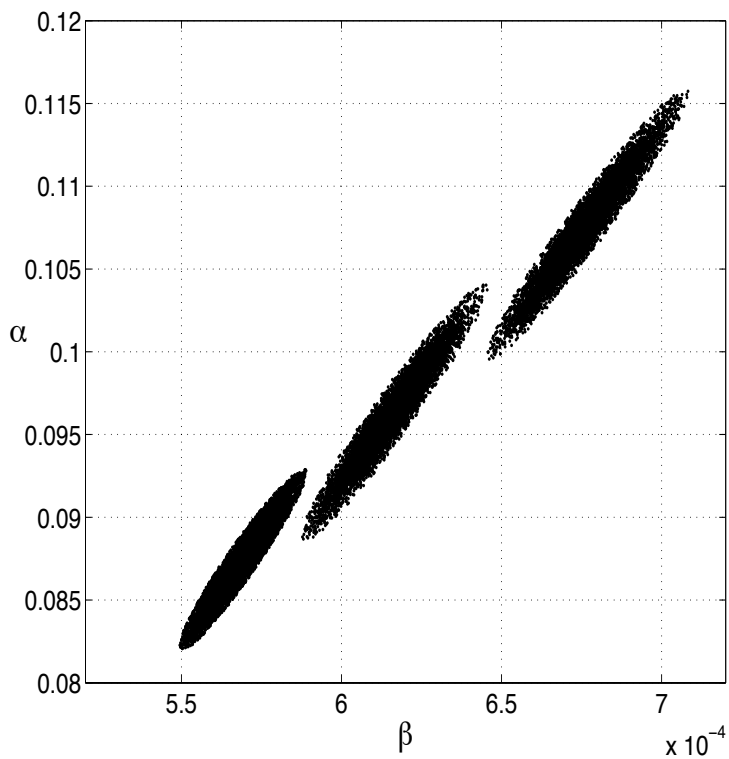

Figure 3. Unconditional confidence regions for the Eyam plague parameters, from bottom left to top right ellipses are for $I(0) = 4, 5$ and $6$.

## 5. DISCUSSION

Assessing estimability typically involves the paradox that parameter estimates are required a-priori to assessing parameter estimability. Additionally, model refinements and re-parameterization require re-running analysis many times with different model variants. Our proposed ANOVA test simplifies this process considerably because estimability of functions of posterior samples can be assessed in the ANOVA test. The ANOVA test results do not depend on parameter scaling, arbitrarily chosen thresholds, or explicit specification of the nature of the relationships between parameters. When inestimble parameters are found, plots of the posterior distributions, may reveal relationships between parameters that can lead to suitable model re-parameterization, further inquiry, or refined model formulation.

Along with introducing a test for estimability, we illustrate the use of DC techniques to conduct statistical inference for partially and imperfectly observed dynamic systems. As demonstrated, one can use DC to conduct inferential procedures such as likelihood ratio testing, profile likelihood inference, and both conditional and unconditional confidence regions with a mixture of continuous and discrete parameters while assessing estimability. The ANOVA test compliments DC analysis by removing ambiguity concerning whether enough clones have been used. Although the methods were applied to dynamic system models, the DC and ANOVA test can be used more generally to assess identifiability and to obtain point and interval estimates in any situation for which the MLE is appropriate.

An additional benefit of this work is that no new specialized programs are needed to implement DC or the ANOVA estimability test. Standard freely available MCMC programs can readily implement DC. In some cases posterior sampling methods such as weighted resampling (Rubin 1988) or importance sampling with optimization (Raftery and Bao 2010) may be required for DC implementation, while more specialized MCMC methods specific to dynamic systems (eg. Campbell and Steele 2011, Calderhead and Girolami 2009) may help with sampling in the presence of difficult posterior topology. Since the estimability results examine the effect of the prior on the MLE, one must choose priors which are far apart, and as a consequence, one may need to use very large numbers of clones to ensure the likelihood offsets the impact of the priors. In the provided examples, cloning amounted to using a large exponent on the likelihood and consequently, there is no real cost to large numbers of clones. However with stochastic differential equation models, $x(t,\varphi,x_0)$ has a distribution that will require sampling in such a way that increasing the number of clones increases the computational load and memory requirements of the MCMC runs. Efficient implementation for stochastic differential equation models is the subject of future research.

When the likelihood is available analytically and the Fisher Information is easy to compute, the proposed methods are more computationally intensive than necessary. Since our goal is maximum likelihood estimation one may consider optimizing the likelihood rather than using MCMC methods. When the likelihood is not available analytically, as is the case in our examples, the hessian of the log likelihood and obtaining it's inverse are subject to numerical instabilities such that lack of estimability may be confounded with numerically rank deficient matrices.

One note of caution for the test for a significant prior effect to assess identifiability is that the priors must be chosen to be far apart, however 'far apart' in the parameter space may not be 'far apart' in the response surface, and the distinction is difficult to assess. Theoretically one could chose priors that are equally distant to a likelihood ridge so as to produce identical posterior specifications when truncated onto the manifold. This result would be the equivalent of a type 2 error in the test for a prior effect. To reduce the probability of such occurrences one could explore improved experimental design strategies for selecting priors, increase the number of priors to reduce the probability of the problem, or choose priors sequentially to be near and far from the posterior parameter estimates.

With any optimization routine, it is good practice to perform estimation from a variety of starting points. In fact MCMC practitioners routinely attempt different independent runs, when applying the Gelman-

Rubin (1992) diagnostic. The ANOVA diagnostic also requires multiple MCMC attempts, yet unlike with the Gelman-Rubin diagnostic our MCMC runs may be all targeting different posterior distributions, depending on the clone and prior effects.

ACKNOWLEDGEMENTS:

We would like to thank the associate editor and two anonymous reviewers for their insightful comments.